\documentclass[cameraready]{Interspeech}

\title{An Approach to Simultaneous Acquisition of Real-Time MRI Video, EEG, and Surface EMG for Articulatory, Brain, and Muscle Activity During Speech Production}

\author[affiliation={1,2}, equalcontribution, orcid=0009-0009-3033-2042]{Jihwan}{Lee}
\author[affiliation={2}, equalcontribution, orcid=0000-0003-0057-4402]{Parsa}{Razmara}
\author[affiliation={1,2}, equalcontribution, orcid=0009-0000-1829-2265]{Kevin}{Huang}
\author[affiliation={1,3}, orcid=0009-0000-1495-4487]{Sean}{Foley}
\author[affiliation={1,2}, orcid=0009-0002-0056-8336]{Aditya}{Kommineni}
\author[affiliation={3}, orcid=0009-0001-4357-0092]{Haley}{Hsu}
\author[affiliation={2}, orcid=0000-0001-5415-2040]{Woojae}{Jeong}
\author[affiliation={2}, orcid=0000-0002-0685-5160]{Prakash}{Kumar}
\author[affiliation={1,2}, orcid=0000-0002-1387-5418]{Xuan}{Shi}
\author[affiliation={1}, orcid=0000-0003-1323-049X]{Yoonjeong}{Lee}
\author[affiliation={1,2}, orcid=0000-0002-2053-9068]{Tiantian}{Feng}
\author[affiliation={2}, orcid=0000-0003-0774-067X]{Takfarinas}{Medani}
\author[affiliation={2}, orcid=0000-0002-8559-4404]{Ye}{Tian}
\author[affiliation={1,2}, orcid=0000-0001-5806-3053]{Sudarsana Reddy}{Kadiri}
\author[affiliation={2}, orcid=0000-0001-5735-3550]{Krishna S.}{Nayak}
\author[affiliation={3}, orcid=0000-0003-3319-5871]{Dani}{Byrd}
\author[affiliation={3}, orcid=0000-0002-5473-4216]{Louis}{Goldstein}
\author[affiliation={2}, orcid=0000-0002-7278-5471]{Richard M.}{Leahy}
\author[affiliation={1,2,3}, orcid=0000-0002-1052-6204]{Shrikanth}{Narayanan}

\address{
    $^1$ Signal Analysis and Interpretation Laboratory, University of Southern California\\
    $^2$ Ming Hsieh Dept. of Electrical and Computer Engineering, University of Southern California\\
    $^3$ Dept. of Linguistics, University of Southern California
}

\email{jihwan@usc.edu}

\keywords{articulation, real-time MRI, EEG, EMG, speech neurophysiology, brain-computer interface (BCI)} 

\usepackage{comment}
\usepackage{tabularx} 
\usepackage{booktabs} 
\usepackage{subcaption}
\usepackage{cite}

\begin{document}

\maketitle
\vspace{-5mm}
\begin{abstract}
    \vspace{-3mm}
Speech production is a complex process spanning neural planning, motor control, muscle activation, and articulatory kinematics. While the acoustic speech signal is the most accessible product of the speech production act, it does not directly reveal its causal neurophysiological substrates. We present the first simultaneous acquisition of real-time (dynamic) MRI, EEG, and surface EMG, capturing several key aspects of the speech production chain: brain signals, muscle activations, and articulatory movements. This multimodal acquisition paradigm presents substantial technical challenges, including MRI-induced electromagnetic interference and myogenic artifacts. To mitigate these, we introduce an artifact suppression pipeline tailored to this tri-modal setting. Once fully developed, this framework is poised to offer an unprecedented window into speech neuroscience and insights leading to brain-computer interface advances. The source code and data are available\footnote{\url{github.com/lee-jhwn/multimodal-speech-biosignals}}.
\end{abstract}

\vspace{-3mm}
\section{Introduction}
\vspace{-2mm}
Speech production is a complex activity, originating from intricate neurocognitive planning and culminating in coordinated neuromuscular action and resulting aerodynamic and acoustic modulation.
While the acoustic output is the most accessible modality for speech production analysis, it is an end product of a complex chain of causal events. To fully understand the mechanisms of spoken language production and build robust biosignal-to-speech systems, it is beneficial to look beyond the audio waveform and examine the causal physiological and neural substrates underlying it.

The physical realization of speech acoustics is governed by coordinated articulatory movements of the vocal tract~\cite{browman1992articulatory}.
Several tools to directly observe articulatory kinematics, such as real-time magnetic resonance imaging (rtMRI) or electromagnetic articulography (EMA), provide useful resources to understand the relationship between articulatory and acoustic outputs~\cite{narayanan2014real, huang25h_interspeech, lim2021multispeaker, usc-lss, foley2025towards, shi2024direct, lee2026arti6sixdimensionalarticulatoryspeech, cho2024coding, ghosh2010generalized, lammert2013statistical, otani2023speech, sorensen2017database, lim2024speech, toutios2019advances, yue2024towards, park2026interpretablemodelingarticulatorytemporal}.


Preceding the articulatory movements is the electrical activation of articulatory muscles, some of which (e.g., orofacial) can be captured by surface electromyography (EMG).
EMG has been used to study the timing of motor recruitment and to \mbox{explore} alternative communication tools to those with speech motor disorders,
such as in silent speech interfaces~\cite{lee25d_interspeech, gaddy2022voicing, schultz2010modeling, diener2018, jou2006towards, scheck-2023, ren-2024, sualiheen2025emgvox, wu24k_interspeech, gowda2025emg2speech}.

At the apex of this speech production hierarchy lies the brain, where semantic intent is translated into physical messages via motor control of the vocal tract. Electroencephalography (EEG) provides a non-invasive approach to measure underlying brain activity with high temporal resolution, which is essential for tracking the rapid dynamics of speech production.
EEG has been widely utilized to investigate the neural basis of language and speech representation and the spatiotemporal dynamics associated with speech processing
critical to brain-computer interfaces (BCIs) or brain-to-speech decoding systems~\cite{puffay2023relating, fesde, fesde2,imaginedspeech23, vlaai, xu2024convconcatnet, bras24_is, meta-paper, shukla2025survey, huang2025ccsumsp, ma2025fully, he2025vocalmind, sato2024scaling, moreira2025open, kurteff2023speaker, avramidis2025neuralcodecsbiosignaltokenizers}.

To investigate the interconnection among brain, muscle activity, and articulation,
previous studies have attempted to combine acquisition of these modalities.
Lezcano et al. report methods for synchronization of EMA and EMG during speech production~\cite{emg-ema-sync-2023}, while Friedrichs et al. explore temporal co-registration of EEG and EMA~\cite{friedrichs24_interspeech}.
Anastasopoulou et al. probe concurrent recording of magnetoencephalography and magneto-articulography to map between brain activity and speech kinematics~\cite{anastasopoulou2022speech}.
These studies, however, are limited to a subset of these modalities and also do not provide the full view of the moving vocal tract during speech production.



In this study, we investigate the technical feasibility of simultaneously acquiring rtMRI, EEG, and surface EMG during speech production. This multimodal biosignal acquisition allows for the concurrent dynamic observation of the major components of speech production chain: brain activation via EEG, orofacial muscle activation via EMG, and articulatory kinematics via rtMRI. To the best of our knowledge, this represents the first attempt to synchronously record these three biosignal modalities during speech production.



The simultaneous acquisition of these modalities presents three primary technical challenges: gradient-induced electromagnetic artifacts, cardiac pulse artifacts, and myogenic contamination. First, rapid switching of magnetic field gradients during MRI acquisition induces large voltage transients in EEG and EMG leads through electromagnetic induction. Second, cardiac-driven motion within the static magnetic field generates the quasi-periodic pulse-related artifact. Third, speech production introduces substantial myogenic contamination, including head and jaw motion that perturbs electrode contacts.
To address these signal contamination challenges, we implement a multi-stage denoising pipeline. Our preliminary results demonstrate attenuation of these artifacts.

Such multimodal data hold utility for advancing both speech science and technology development. From a scientific perspective, it offers a window into the overall neurophysiology of speech production, allowing researchers to investigate the spatiotemporal dynamics from speech planning through motor execution to the resulting articulatory movements.
In the BCI domain, these data provide articulatory trajectories even in the absence of acoustic signal, useful for mapping neural activity to articulatory movements in silent speakers. rtMRI also serves as an observation tool for imagined speech studies, allowing unconscious micro-articulatory movements to be monitored that could influence BCI performance.

\vspace{-3mm}
\section{Methods}
\vspace{-3mm}
\begin{figure}[t]
    \centering
    \includegraphics[width=1.0\linewidth]{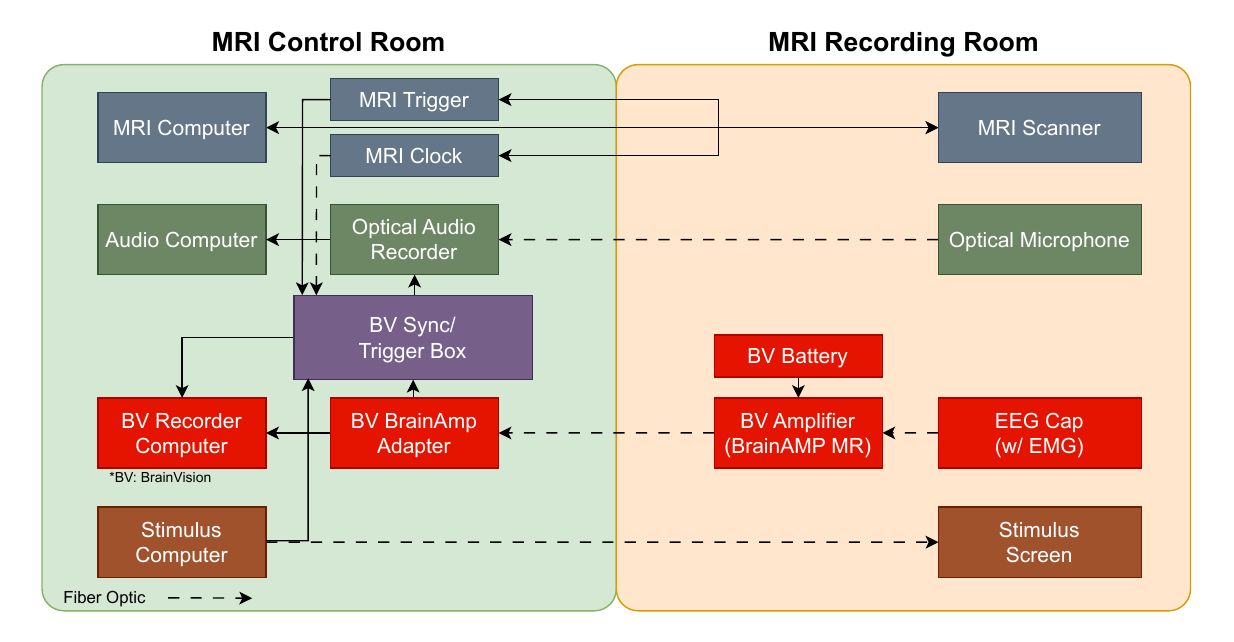}
    \vspace{-9mm}
    \caption{Experimental setup for simultaneous acquisition of rtMRI, EEG, and EMG, along with audio.}
    \label{fig:illus}
    \vspace{-7mm}
\end{figure}
\subsection{Apparatus}
\vspace{-1mm}
The rtMRI vocal tract images are acquired in a 0.55T MRI via a custom 8-channel upper airway coil using a spiral bSSFP sequence with TR set to $5.05$ ms ($99$ fps) with concurrent audio recording~\cite{lim2024speech, kumar24b_interspeech} using an optical microphone\footnote{Optoacoustics Ltd., Moshav Mazor, Israel}. The electrophysiology signals are recorded using an MR-compatible BrainVision system\footnote{Brain Products GmbH, Germany}. Signals are sampled at 5 kHz, synchronized with the MRI clock via fiber-optic trigger.
The 0.55T field strength improves compatibility with electrophysiological recordings ~\cite{razmara2026feasibilitysimultaneouseegfmri055, razmara2025feasibilitystudytaskbasedfmri}. 
The device has 16 electrodes, 9 EEG (C3, C4, F3, F4, FPz, O1, O2, M1, and M2), two EOG, three EMG, and one ECG electrodes.
The three surface EMG electrodes are placed on the following locations of the face, near the muscle areas correlated with articulatory movements~\cite{lee25d_interspeech}: chin corner, underneath the chin, and next to Adam's apple.
The EOG (electrooculography) and ECG (electrocardiogram) electrodes capture the corneo-retinal standing potential and cardio signals, respectively, and are used for artifact removal in EEG signals.

\vspace{-2mm}
\subsection{Experimental design}
\vspace{-1mm}
The pilot study recording consists of data from one male American English native speaker, aged in his thirties, with no visual or hearing impairments. There are three different types of speech production tasks: fully phonated, silent, and imagined speech production, each recorded under two different magnetic field conditions (inside and outside the scanner). This paper primarily focuses on the fully phonated task.
For the phonated task, the subject was instructed to fully produce the stimuli, and for the silent speech task, the subject was instructed to do the same without vocalizing. These tasks are repeated 12 times for each magnetic condition.
The imagined speech production was included as an additional study, and consists of a subset of the full stimulus set with only 2 runs. In this task, the subject was asked to imagine producing each stimulus without any articulatory movements.

The stimulus set consists of 18 disyllabic nonce words with a VCV structure, where the vowel is /a/, a selected subset from \cite{lim2021multispeaker}: [apa, ata, aka, asa, asha, ala, afa, ara, aha, awa, aya, aba, ada, aga, atha, ama, ana, ava]. Prior to each unique speech production task, a 60-second resting-state period was recorded during which a blank screen was presented.
Each trial followed a fixed sequence: (1) a variable inter-trial interval (jittered) consisting of a black screen with a duration randomized between 0.5, 0.75, and 1.0 s to prevent time-locking of stimulus onset to the MRI volume acquisition cycle; (2) a white fixation cross presented in the center of the black background for 0.5 s; (3) the target nonce word presented for 1.0 s, during which the subject was instructed not to produce the word yet; (4) a Go signal (fixation cross) lasting 2.0 s, signaling the onset of speech production; and (5) a final blank screen for 0.5 s before the next trial. The subject was asked not to produce the word in step (3) to minimize visual interference in neural signals.
The stimulus order was randomized.
The study was approved by the Institutional Review Board of our institution, and the participant provided written informed consent prior to participation.





\begin{figure}[t]
    \centering
    \includegraphics[width=1.0\linewidth]{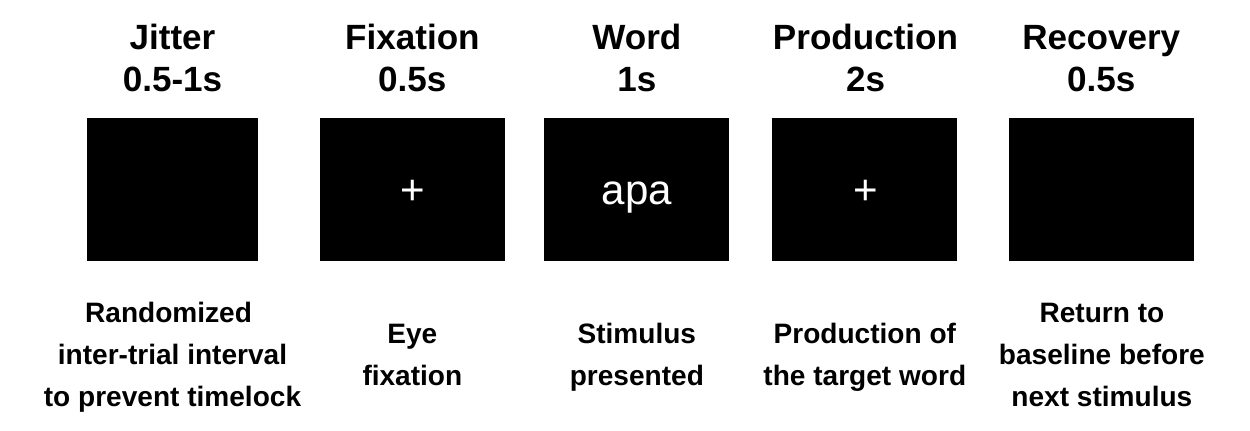}
    \vspace{-8mm}
    \caption{Stimulus presentation protocol.}
    \label{fig:illus}
    \vspace{-8mm}
\end{figure}

\vspace{-2mm}
\subsection{Magnetic artifact removal}
\vspace{-1mm}

Simultaneous EEG recording during rtMRI is contaminated by gradient switching artifacts (GA) and ballistocardiogram (BCG) artifacts. We implement a two-stage correction procedure: (1) gradient artifact correction followed by (2) pulse (BCG) artifact correction, both using template-based subtraction adapted from simultaneous EEG-fMRI preprocessing frameworks \cite{allen2000method, allen1998identification, warbrick2022simultaneous, razmara2026feasibilitysimultaneouseegfmri055}. 
Magnetic artifact correction is performed in BrainVision Analyzer 2 \cite{brainvision_analyzer_2023}. Data inspection and visualization are conducted in Brainstorm \cite{tadel2011brainstorm}.

\vspace{-2mm}
\subsubsection{Gradient artifact correction}
\vspace{-1mm}
Rapid gradient switching during rtMRI induces large-amplitude, periodic voltage transients in EEG leads through electromagnetic induction. Due to the strict periodicity of the gradient waveform, gradient artifact removal is performed using average artifact subtraction~\cite{allen2000method}.
Artifact onset markers are detected from the EEG using a gradient-threshold method, identifying the start of each repeating artifact pattern aligned with the acquisition cycle. For each repetition, a local artifact template is constructed by averaging a centered sliding window of repetitions, then subtracted from the data~\cite{allen2000method}. The sliding window accommodates gradual changes in artifact shape caused by subject motion. Correction parameters follow the default settings of BrainVision Analyzer 2 and are adapted to our rtMRI setup~\cite{brainvision_analyzer_2023, allen2000method}.




\vspace{-2mm}
\subsubsection{Pulse (BCG) artifact correction}
\vspace{-1mm}
After gradient artifact removal, residual cardiac-locked deflections are identified. R-peaks are detected from the simultaneously recorded ECG channel (low-pass filtered at 15 Hz) and verified semi-automatically. An ECG-informed average artifact subtraction approach is applied, in which cardiac-locked segments are averaged within a sliding window and subtracted from the EEG signal \cite{allen1998identification,warbrick2022simultaneous}. This procedure reduces ECG-locked waveform components and attenuates bilaterally mirrored, polarity-reversed patterns across channels, consistent with suppression of the BCG artifact. 




\begin{figure}[t!]
\vspace{-5mm}
    \centering
    \begin{subfigure}[b]{0.49\linewidth}
        \centering
        \includegraphics[width=\linewidth]{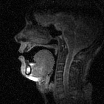} 
        \caption{Without EEG/EMG cap}
        \label{fig:plot1}
    \end{subfigure}
    \hfill 
    \begin{subfigure}[b]{0.49\linewidth}
        \centering
        \includegraphics[width=\linewidth]{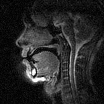} 
        \caption{With EEG/EMG cap}
        \label{fig:plot2}
    \end{subfigure}
    \vspace{-3mm}
    \caption{\textbf{Comparison of rtMRI videos with and without the EEG/EMG device.} We observe no significant impact on the articulatory regions of interest.}
    \label{fig:withandwithouteegmri}
    \vspace{-7mm}
\end{figure}

\begin{figure}[t]
    \vspace{-5mm}
    \centering
        \begin{subfigure}[b]{\linewidth}
        \centering
                \caption{Raw EEG recording inside the running scanner.}
        \includegraphics[width=\linewidth]{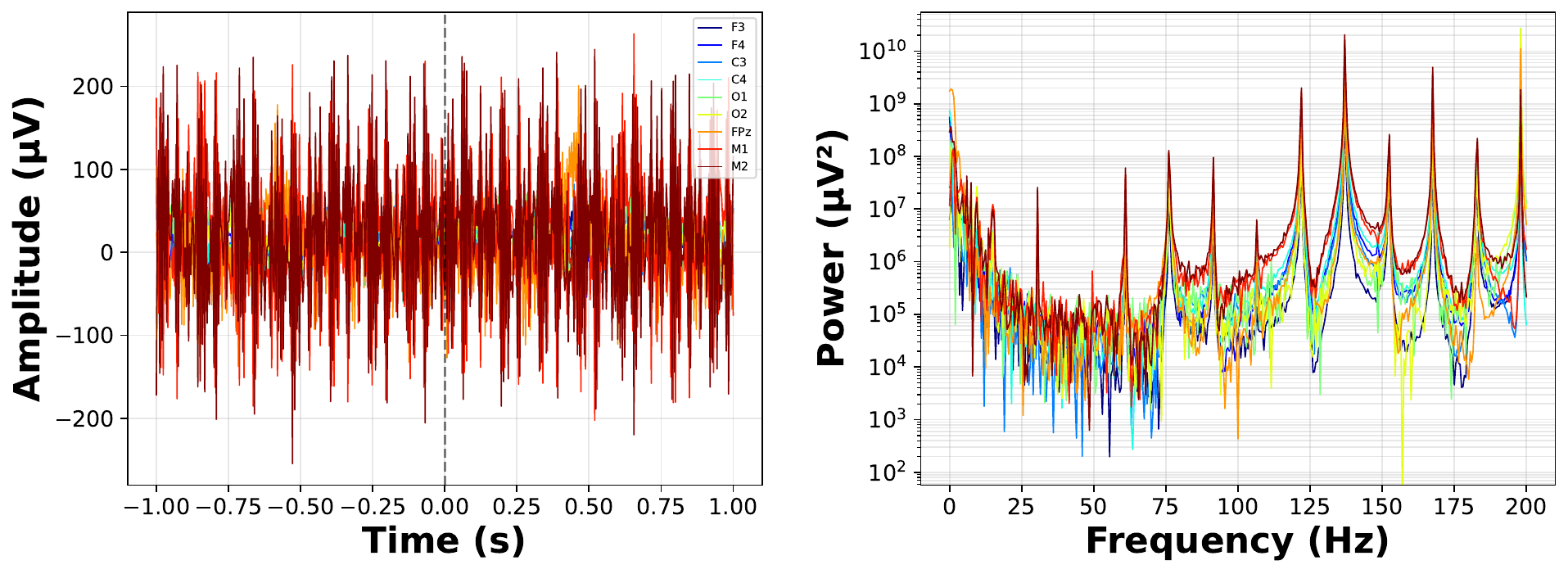} 
        \label{fig:raweeg}
        \vspace{-6mm}
    \end{subfigure}
        \begin{subfigure}[b]{\linewidth}
        \centering
                \caption{Inside scanner condition after magnetic artifact correction.}
        \includegraphics[width=\linewidth]{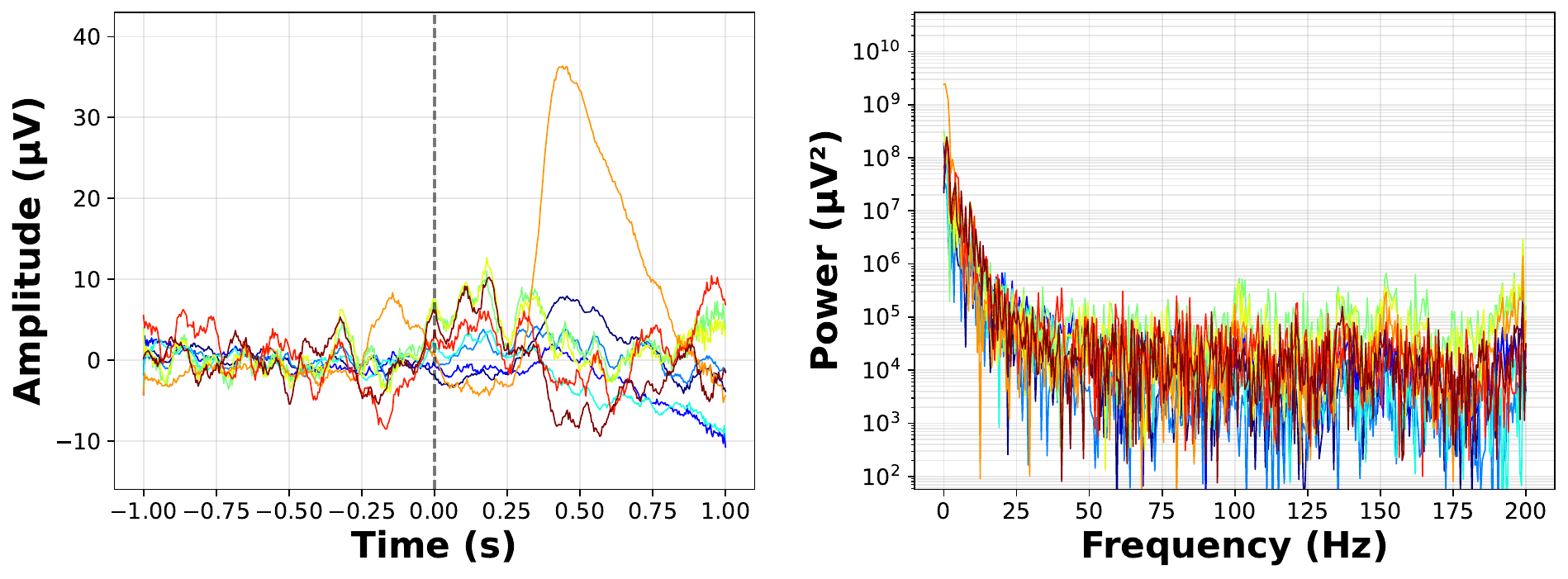} 
        \label{fig:aftermag}
            \vspace{-6mm}
    \end{subfigure}
        \begin{subfigure}[b]{\linewidth}
        \centering
                \caption{Outside scanner reference.}
        \includegraphics[width=\linewidth]{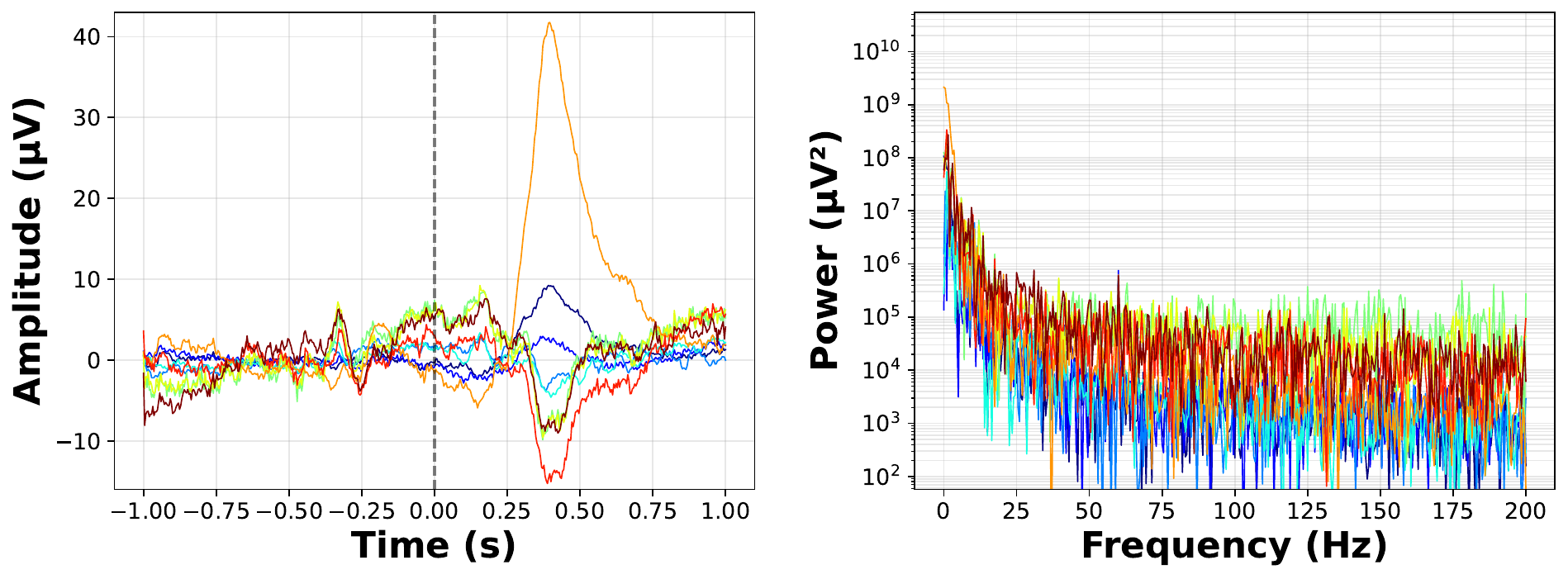} 
        \label{fig:outsideeeg}
    \end{subfigure}
    \vspace{-10mm}
    \caption{\textbf{Magnetic artifact correction around stimulus onset.}
    Raw EEG recording inside the scanner (Fig.~\ref{fig:raweeg}) shows large-amplitude periodic transients by gradient switching, which is substantially suppressed after the correction (Fig.~\ref{fig:aftermag}), exhibiting temporal and spectral characteristics comparable to the reference (Fig.~\ref{fig:outsideeeg}). Each row presents representative time-domain waveforms (left) and corresponding frequency magnitude spectra (right).
    }
    \label{fig:compare_mag}
    \vspace{-7mm}
\end{figure}

\vspace{-2mm}
\subsection{Myogenic and ocular artifact correction}
\vspace{-1mm}
Residual myogenic and ocular artifacts remain in the EEG signal, arising from speech-related facial muscle activity, eye blinks, and eye movements, and may confound analyses by temporally overlapping with task-related neural signals. To suppress these non-neural components while preserving cortical activity, we applied reference-based canonical correlation analysis (CCA) using EMG and EOG channels as references~\cite{de2006canonical, mucarquer2019improving}.

EEG ($9$ channels) and artifact reference signals ($3$ EMG, $2$ EOG) were jointly decomposed using CCA to identify components maximizing correlation between brain and peripheral signals. Components exhibiting strong correlation with the reference channels were classified as artifactual and removed via orthogonal projection of the corresponding subspace. Components were ranked by canonical correlation ($\rho$) with the reference signals. Components with $\rho>0.4$ were rejected, supplemented by visual inspection of component time courses and spatial patterns \cite{de2006canonical, gao2010online}. This resulted in removal of 2 to 5 canonical components per recording. 

Effectiveness is evaluated by: 
(i) suppression of blink-locked frontal artifacts synchronized with EOG activity; 
(ii) attenuation of movement-locked artifacts synchronized with EMG activity during articulation; and 
(iii) visual inspection of scalp topographies for restoration of expected speech production patterns.

\begin{figure*}[t]
\vspace{-5mm}
    \centering
    \begin{subfigure}[b]{0.49\textwidth}
        \centering
        \includegraphics[width=\textwidth]{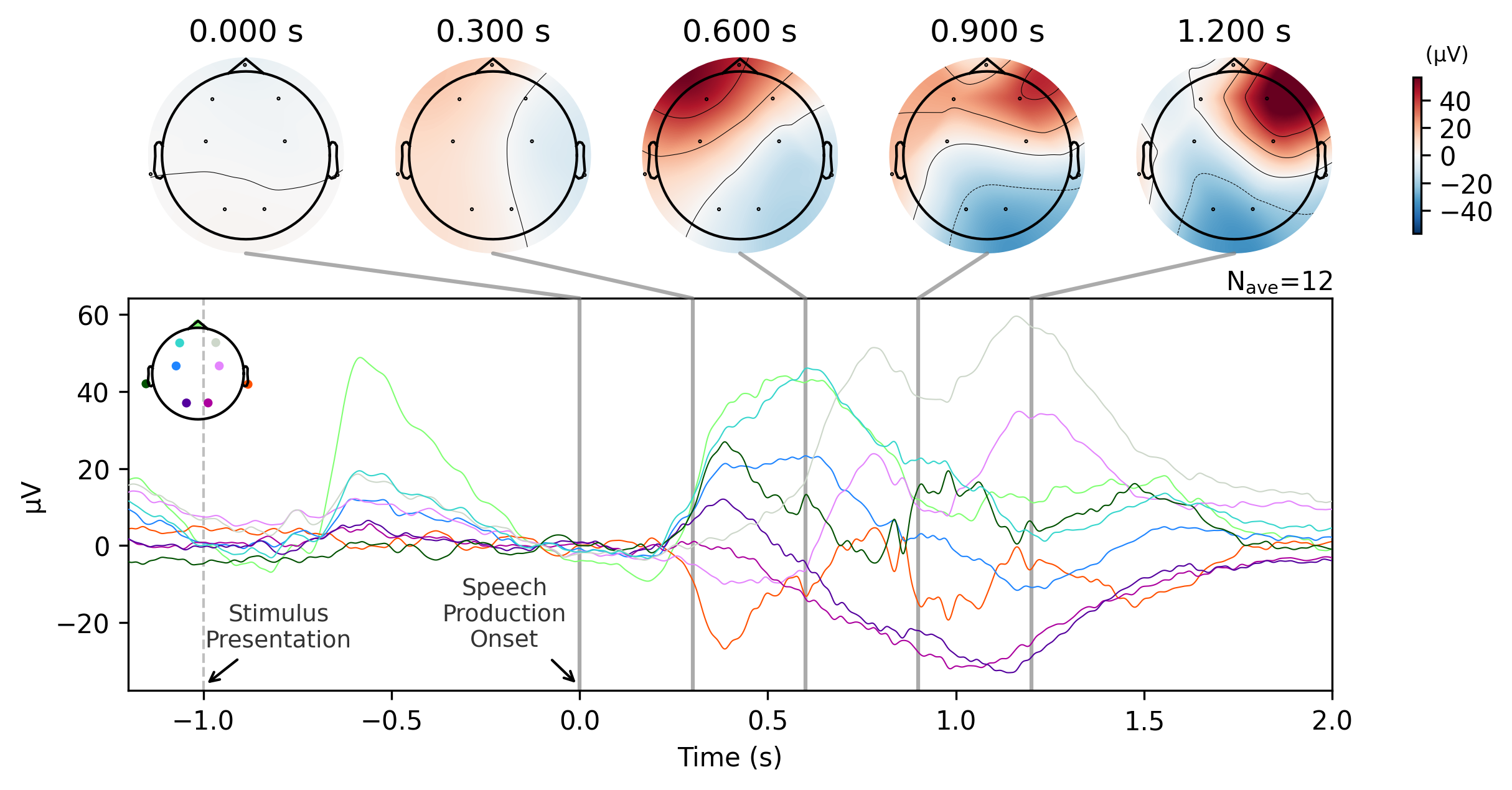}
                        \vspace{-6mm}
        \caption{Before Myogenic and Ocular Artifact Removal}
        \label{fig:musclebefore}
    \end{subfigure}
    \begin{subfigure}[b]{0.49\textwidth}
        \centering
        \includegraphics[width=\textwidth]{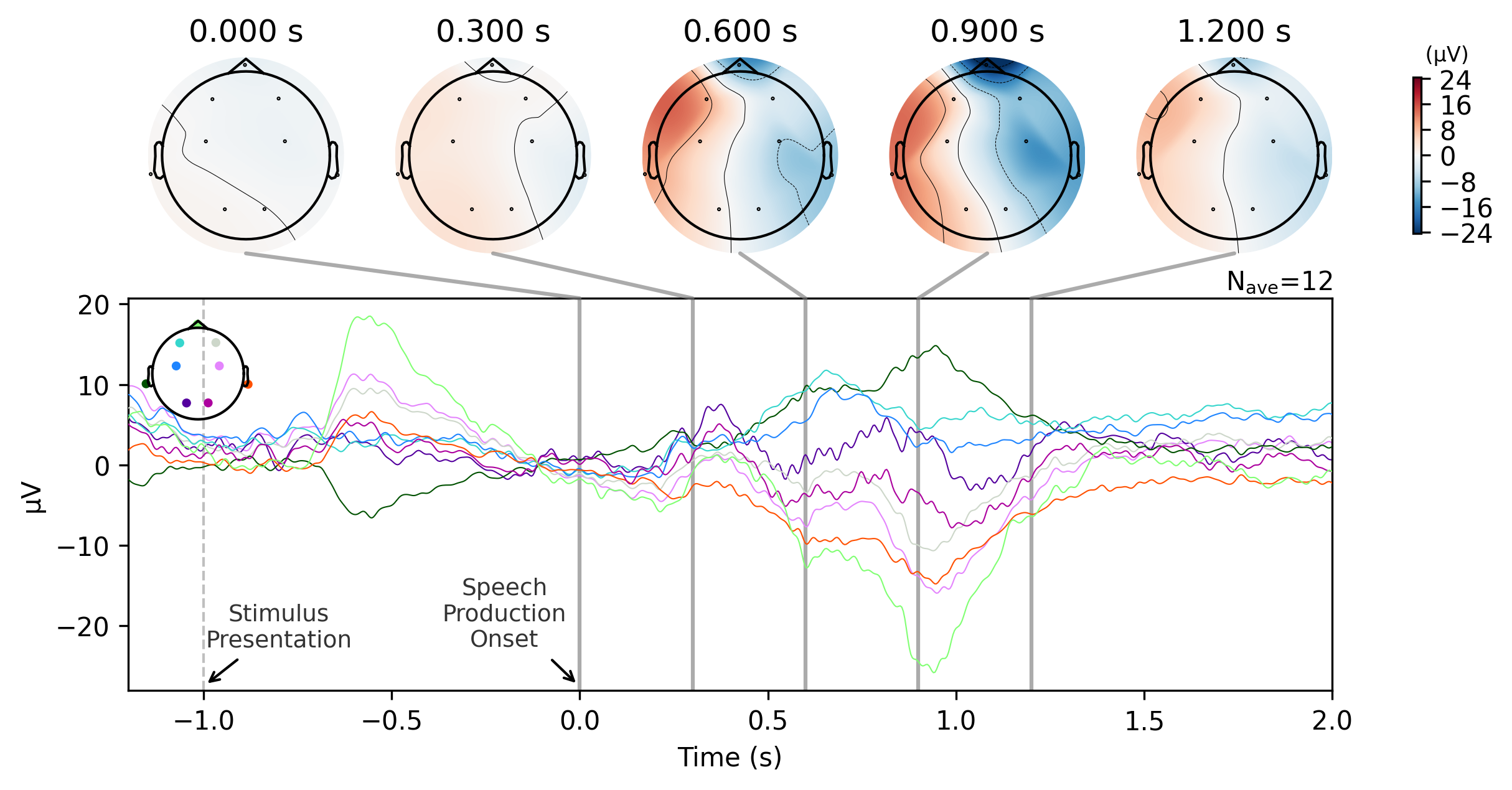}
                \vspace{-6mm}
        \caption{After Myogenic and Ocular Artifact Removal}
        \label{fig:muscleafter}
    \end{subfigure}
    \ifcameraready
    \vspace{-3mm}
    \fi
    \caption{\textbf{Comparison of ERPs and topographies before and after myogenic and ocular artifact removal.} Prior to artifact removal, high-amplitude activity is heavily concentrated in the frontal region, indicative of artifact contamination. After the denoising pipeline, this frontal dominance is substantially attenuated, revealing a distinct lateralization of activity over the left hemisphere, consistent with language processing areas.}
    \label{fig:comparemuscle}
    \ifcameraready
    \vspace{-5mm}
    \fi
\end{figure*}

\vspace{-3mm}
\section{Results and Discussion}
\vspace{-1mm}

\subsection{Temporal alignment}
\vspace{-1mm}
The EEG, EOG, and EMG signals are acquired via a single system and are intrinsically aligned. Given that MRI-audio synchronization was previously established~\cite{lim2024speech, kumar24b_interspeech}, we inspect the MRI-EEG alignment by comparing the total duration of the MRI video to the interval between the scanner’s start and stop triggers in the EEG recording. The average duration difference is $8.3$ ms/s ($\sigma=0.3$ ms/s), within the range of the frame size of the rtMRI videos ($10.1$ ms).

\vspace{-1mm}
\subsection{Influence of EEG/EMG setups on rtMRI videos}
\vspace{-1mm}
All of the elements such as electrodes and cords of the EEG and EMG recording are MRI-compatible, hence there is no or negligible effect on the recorded rtMRI videos. Fig.~\ref{fig:withandwithouteegmri} compares the rtMRI videos with and without the EEG/EMG electrodes, and no significant artifacts from the EEG/EMG electrodes are observed in the rtMRI images.
A region of interest analysis on the tongue region yields an SNR of $10.148\pm0.575$, which is within the bounds of expected SNR reported in \cite{lim2024speech}, indicating no significant quantitative impact on the recorded rtMRI videos from simultaneous use of EEG/EMG electrodes.

\vspace{-2mm}
\subsection{Magnetic artifact correction}
\vspace{-1mm}

We address the effectiveness of the magnetic artifact removal method (GA and BCG correction).
We mainly compare three different conditions: (1) Inside the scanner before any magnetic denoising, (2) Inside the scanner after magnetic denoising, and (3) Outside the scanner (no magnetic noise) as a reference point.

As shown in Fig.~\ref{fig:raweeg}, the raw EEG recordings inside the scanner contain periodic high-amplitude transients and harmonic spectral peaks that are generally not observed in EEG recordings. After magnetic denoising, the high-frequency harmonics are removed (Fig.~\ref{fig:aftermag}).


We also compare the ERPs of two-second windows, one second before and after the onset of stimulus presentation. It shows average temporal correlation of 0.66 ($\sigma=0.17$) between the inside and outside scanner conditions, especially higher correlation at the frontal pole and near the language-production activity (FPz, C3, F3 reaching around 0.82, 0.81, 0.78, respectively). Note that this time window is prior to articulation of words, before myogenic contamination starts.


\vspace{-3mm}
\subsection{Myogenic and ocular artifact correction}
\vspace{-2mm}

To evaluate the efficacy of the myogenic and ocular artifact removal pipeline, we compare ERPs and scalp topographies before and after CCA denoising, as shown in Fig.~\ref{fig:comparemuscle}. Prior to artifact removal (Fig.~\ref{fig:musclebefore}), uncorrected data show severe myogenic and ocular contamination reaching roughly 60~$\mu$V. Following the application of the reference-based CCA pipeline (Fig.~\ref{fig:muscleafter}), the underlying neural signal emerges, exhibiting left-lateralized activation broadly consistent with prior reports of speech-related cortical processing~\cite{hickok2007cortical,friederici2011brain, price2012review}. The correction substantially attenuates the initial contamination, reducing peak amplitudes across most channels to roughly 20~$\mu$V and suppressing widespread myogenic artifacts across cortical frontal channels such as F3 and F4. A localized residual artifact persists at the extreme front of the head (FPz), likely reflecting mechanical perturbation of the electrode during speech-induced head motion. As the frontal pole lies outside primary speech and motor networks, this isolated contamination does not obscure the cortical dynamics of interest.





\ifcameraready
\vspace{-3mm}
\fi
\subsection{Micro-articulatory movements in imagined speech}
\ifcameraready
\vspace{-2mm}
\fi
\begin{figure}[t!]
\vspace{-2mm}
    \centering
    \begin{subfigure}[b]{0.32\linewidth}
        \centering
        \includegraphics[width=\linewidth]{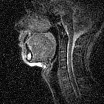} 
        \label{fig:plot1}
    \end{subfigure}
    \begin{subfigure}[b]{0.32\linewidth}
        \centering
        \includegraphics[width=\linewidth]{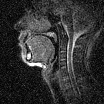} 
        \label{fig:plot2}
    \end{subfigure}
    \begin{subfigure}[b]{0.32\linewidth}
        \centering
        \includegraphics[width=\linewidth]{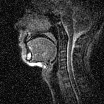} 
        \label{fig:plot2}
    \end{subfigure}
    \vspace{-6mm}
    \caption{\textbf{Micro-articulatory movements during imagined speech.} Snapshots of rtMRI videos during imagined production of /ama/. Micro-articulatory movements are observed, such as the velum movement.}
    \label{fig:imagine-ama}
    \vspace{-7mm}
\end{figure}
We observe micro-articulatory movements during imagined speech production, as shown in Fig.~\ref{fig:imagine-ama}. Although the subject was  instructed to inhibit any articulatory movement, the data suggest that such micro-motor activity persists despite conscious inhibitory efforts, aligning with previous studies~\cite{orpella2022decoding}. Our approach offers an objective means for detecting such subtle articulatory motor activity during imagined speech tasks. Further investigation is required to determine whether these movements can be fully inhibited through training or if they represent a systematic, intrinsic component of the speech-planning process. Regardless, our methodology provides a novel framework for observing the relationship between imagined speech production and micro-motor execution.

\ifcameraready
\vspace{-3mm}
\fi
\section{Future Potential Applications}
\ifcameraready
\vspace{-2mm}
\fi
Once fully developed, this novel multimodal methodology promises advances in two major domains. First, it offers valuable resources to the BCI community, especially for speech decoding from neural or muscle signals. Decoding partially available forms of speech, such as silent or imagined speech, has historically been challenging due to absence of audio ground-truth and the inability to concurrently observe articulator movements. By providing simultaneous articulatory kinematic, muscular, and neural data, this framework establishes a physiological ground truth even when no sound is produced, enabling the training of such robust decoders that map neural or muscle activity directly to articulatory movements.

Secondly, in the speech science domain, this multimodal approach may help uncover the complex mechanisms of speech planning and production. By capturing synchronous multimodal biosignals across the speech production chain, researchers can investigate the spatiotemporal coordination between neural planning, motor execution, and physical articulation. For instance, this data could be pivotal in distinguishing between feedforward motor commands and sensory feedback loops, or in identifying the specific neural breakdowns associated with speech disorders like stuttering or apraxia.

\ifcameraready
\vspace{-3mm}
\fi
\section{Limitation}
\ifcameraready
\vspace{-2mm}
\fi

Despite the promising integration of EEG, rtMRI, and EMG, several limitations remain, including the restricted generalizability of this single-subject pilot study and the technical trade-offs inherent in the hardware. The MRI-compatible EEG electrodes are passive electrodes, and noise level may be higher than active electrodes. Also, the EEG cap is not specifically designed for a study with our type of speech tasks, hence it may not prioritize the capture of speech or language specific regions of the brain, and it certainly has limitations in its spatial resolution. Additionally, the limited EMG montage leaves some residual artifacts. Finally, there may have been auditory and visual consequences on neural activities due to the scanner noise and visually presented orthographic stimuli.


\ifcameraready
\vspace{-3mm}
\fi
\section{Conclusion and Future Work}
\ifcameraready
\vspace{-2mm}
\fi
This study demonstrates the technical feasibility of simultaneously acquiring real-time MRI, EEG, and surface EMG during speech production. This novel trimodal framework takes initial steps to bridge the gaps connecting neural activity, motor execution, and articulatory movements, offering a comprehensive physiological view spanning various aspects of speech production. These capabilities pave the way for developing more robust, physiologically grounded BCIs and can lead to new  insights into the complex neural mechanisms driving human speech.

We anticipate that the proposed acquisition and denoising pipeline is generalizable to various MRI scanner types and high-density EEG devices. Future work will focus on scaling this protocol to a larger cohort of subjects and employing a higher spatial resolution EEG device. Additionally, we aim to investigate more linguistically complex tasks to fully exploit the multimodal richness of this setup.


\section{Acknowledgments}
\ifcameraready
\noindent This work was supported by the US National Science Foundation (IIS-2311676, BCS-2240349) and the US National Institute of Biomedical Imaging and Bioengineering of the US National Institutes of Health (R01EB026299).
\else 
[Hidden for double-blind submission.]
\fi

\section{Generative AI Use Disclosure}
A generative AI tool (Google Gemini Pro) was used for editing text, but not for producing any significant content of the manuscript.

\bibliographystyle{IEEEtran}
\bibliography{mybib}

\end{document}